\documentclass{IEEEmce}

\usepackage{cite}
\usepackage{tikz}
\usetikzlibrary{quantikz2}
\usepackage{amsmath,amssymb,amsfonts}
\usepackage{algorithmic}
\usepackage{graphicx}
\usepackage{subcaption}
\usepackage{textcomp}
\usepackage[table]{xcolor}
\usepackage{soul}
\usepackage{braket}
\usepackage{array}
\usepackage{makecell}
\usepackage{booktabs}
\usepackage{listings}
\usepackage{xparse}
\usepackage{comment}
\usepackage{balance}
\usepackage{hyperref}

\hypersetup{
    colorlinks=true,
    linkcolor=blue,
    filecolor=blue,
    citecolor=blue,
    urlcolor=blue
}

\sethlcolor{yellow}

\renewcommand{\arraystretch}{2}
\newcommand{\etal}{\textit{et al.~}}

\newcommand{\eg}{\textit{e.g.}}
\newcommand{\viz}{\textit{viz.}}
\newcommand{\via}{\textit{via~}}

\jname{IEEE Consumer Electronics Magazine}

\begin{document}

\sptitle{Quantum Sensing in Consumer Electronics}


\title{Quantum Sensing Applications in Consumer Electronics: Emerging Methods and Near-Term Use Cases}


\author{Sounak Bhowmik, Himanshu Thapliyal}


\affil{Southern Methodist University}

\markboth{Bhowmik and Thapliyal}{Quantum Sensing in Consumer Electronics}


\begin{abstract} 
Quantum sensing is emerging as a practical means of improving sensing capabilities in consumer electronics, particularly where size, power, and noise constrain classical approaches. In this paper, we reviewed the near-term applicability of quantum sensing using a platform-and-application framework. Within this framework, we outlined the general sensing protocol and discussed the working of the four key quantum sensing platforms relevant to consumer electronics: nitrogen-vacancy (NV) centers, atomic vapor cells, Rydberg atomic sensors, and photonic quantum sensors. Through case studies involving navigation and positioning, healthcare and wearables, radio-frequency (RF) and wireless systems, the Internet of Things (IoT), and imaging, we examined the tangible engineering advantages offered by quantum sensing. Rather than replacing devices outright, quantum sensing is more likely to earn its place in the near-term by upgrading specific measurement components within hybrid classical-quantum systems. Right now, the clearest opportunities are in weak-field detection, compact RF reception, and sensing that remains robust in noise.

\end{abstract}

\maketitle

\enlargethispage{10pt}

\begin{figure*}[htb]
    \centering
    \includegraphics[width=2\columnwidth]{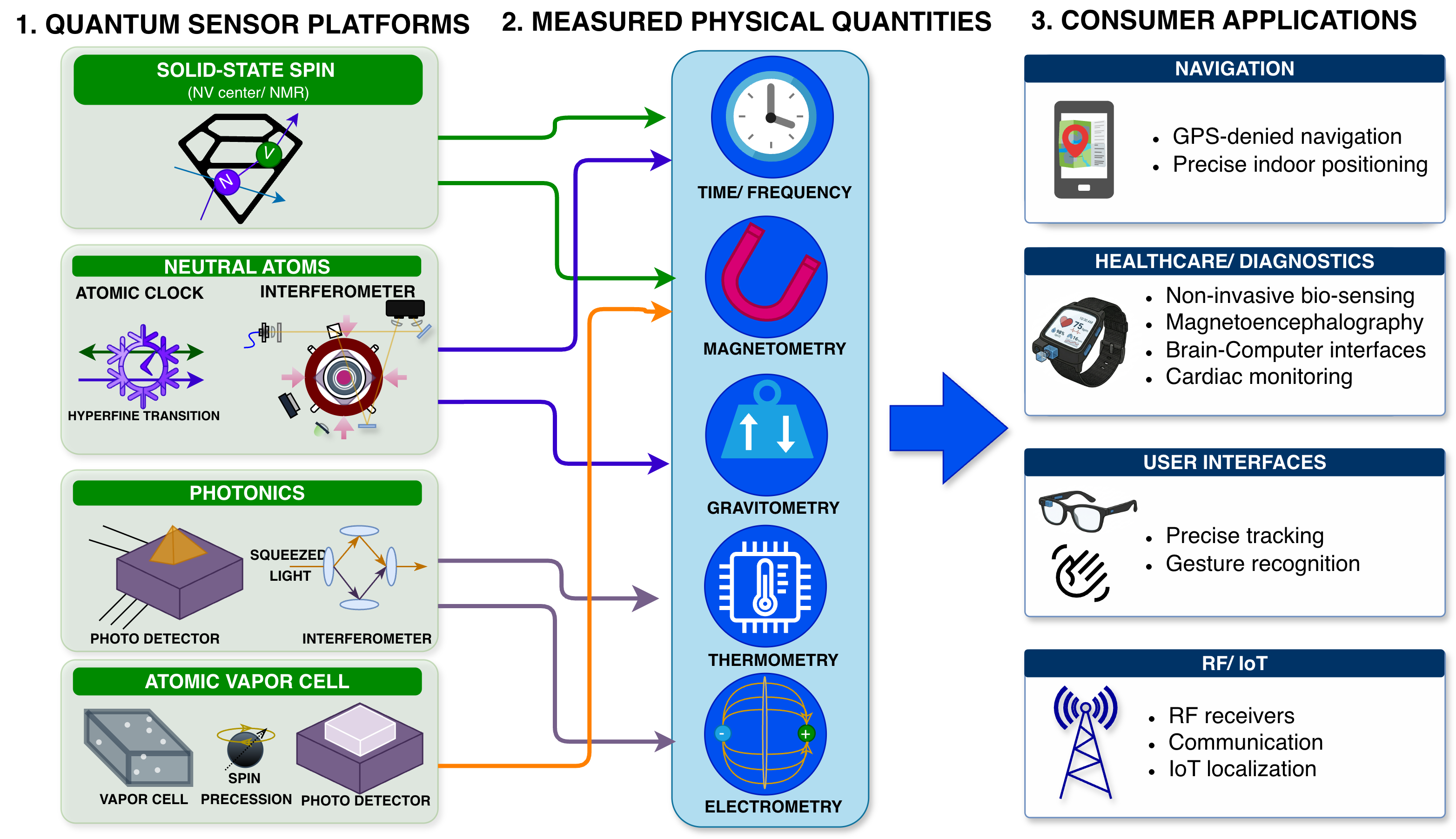}
    \caption{Overview of the quantum sensing landscape for consumer electronics.
    \footnotesize{\\ The quantum-mechanical platforms couple to physical quantities such as magnetic and electric fields, time and frequency, temperature, and gravity, thereby enabling applications in navigation, healthcare, imaging and user interfaces, RF and wireless systems, and the IoT. The key takeaway is that each quantum sensing platform maps most naturally onto the application blocks that depend on the physical quantities it is most sensitive to. For a clearer overview of the generic state-preparation, signal-interaction, and readout sequence involved in a holistic quantum sensing process, refer to Fig~\ref{fig: pipeline}.
}}
    \label{fig: overview}
\end{figure*}

\section{$\mathrm{I}$ Introduction}
\chapterinitial{Quantum sensing} applications are built around highly sensitive quantum states that respond strongly to external perturbations~\cite{QS_Degen}. This sensitivity can be used to detect weak signals that are difficult to resolve using conventional approaches~\cite{QS_Degen}, in a properly engineered and controlled environment. This capability is immensely useful in consumer electronics, where sensitivity, precision, power consumption, form factor, and operational stability impose severe design constraints~\cite{humble2018consumer}, also highlighted at IEEE Quantum Week, 2024~\cite{sotelo20243rd}. Therefore, the practical opportunity in consumer electronics lies not only in achieving higher sensitivity but also in enabling a different sensing architecture. In various applications, a quantum object can serve as a specialized front end or coprocessor embedded within a larger classical system, handling the one measurement task it does the best.

Recent results indicate that this transition is technologically credible. For example, atomic vapor magnetometers have demonstrated that microfabricated vapor cells can achieve subpicotesla ($\text{pT}$) and even femtotesla ($\text{fT}$) level of sensitivity using relatively compact optical configurations~\cite{QS_Degen,shah2007subpicotesla}. This level of performance makes them useful in portable magnetic-sensing modules. Rydberg atomic sensors have demonstrated that atoms can directly transduce radio-frequency (RF) fields and enable optical readout. In certain receiver architectures, this capability reduces dependence on conventional antenna, mixer, and modulator front ends~\cite{fancher2021rydberg,kubler2010coherent}. Studies of thermal-vapor microcells further indicate that useful coherence can be retained under micrometer-scale confinement when the atomic state and cell material are carefully selected~\cite{kubler2010coherent}. Photonic quantum sensing also shows promising results in weak-signal detection and operation at low photon numbers~\cite{pirandola2018advancesinPhotonicQS}. Photonic sensing approaches have also proven reliable where noise tolerance is critical, such as imaging, depth sensing, and biomedical readout~\cite{QS_Degen,pirandola2018advancesinPhotonicQS}.

Collectively, these developments have shifted the key question. It is no longer about whether quantum sensing is practically possible, but about where it delivers a system-level advantage. Moreover, engineers are interested in finding ways to reliably integrate it into practical hardware~\cite{QS_Degen}. In Fig~\ref{fig: overview}, we summarized the relationships among sensing platforms, measured physical quantities, and consumer electronics application domains, and presented a detailed generic sensing workflow in Fig.~\ref{fig: pipeline} that provides a deeper understanding of the underlying mechanisms.

Although prior reviews have examined the underlying principles of quantum sensing, individual sensing platforms, and their broader scientific and industrial applications~\cite{QS_Degen,pirandola2018advancesinPhotonicQS}, this review focuses specifically on the applications in consumer electronics. Here, we examined how quantum sensing technologies can be integrated into practical consumer devices by considering system-level design, near-term application opportunities, and engineering constraints. In this paper, we aim to evaluate where quantum sensors will most likely offer meaningful advantages over conventional sensing technologies in real-world consumer electronics applications.

Moreover, this paper also lays out the conditions under which quantum sensing can offer a meaningful engineering advantage, while satisfying the usual constraints on form factor, power consumption, noise, and operating environment. We focus on scenarios in which quantum sensing can be beneficially integrated into an otherwise classical architecture to enhance critical sensing functions.

The main contributions of this paper are as follows:
\begin{itemize}
    \item We present a generalized, consumer electronics-oriented sensing pipeline and framework that connects key quantum sensing steps, \viz, quantum-state preparation, signal interaction, measurement, and classical post-processing, to consumer system interfaces under practical integration constraints.
    \item We review the operating principles of four quantum sensing platforms of particular relevance to consumer electronics: nitrogen-vacancy (NV) centers in diamond, atomic vapor cells, Rydberg atomic sensors, and photonic quantum sensors.
    \item We present case studies spanning navigation and positioning, healthcare and wearables, RF and wireless systems, the Internet of Things (IoT), and imaging to assess the near-term application space of these platforms.
\end{itemize}

The remainder of this paper is organized as follows. 
Section-II defines the quantum sensing protocol. 
Section-III reviews the sensing platforms that are relevant to understanding the case studies. 
Section-IV describes a general workflow for integrating quantum sensors into consumer electronics systems. 
Section-V examines representative case studies in relevant fields of consumer electronics, such as navigation and positioning, healthcare and wearables, RF and wireless systems, IoT, and imaging. 
Finally, Section-VI summarizes the key challenges and outlines possible future directions.

\section{$\mathrm{II}$ The Quantum Sensing Protocol} \label{sec: SensingProtocol}

Quantum sensing refers to the use of quantum-mechanical systems to measure physical quantities. Following the taxonomy summarized by Degen \etal~\cite{QS_Degen}, quantum sensing may involve one or more of the following elements:
\begin{enumerate}
    \item \textbf{Quantum objects}: Measurements performed using systems with quantized energy levels, such as spin qubits, superconducting circuits, trapped ions, neutral atoms, or photons.
    
    \item \textbf{Quantum coherence}: Measurements performed using spatial (\eg, a particle traversing two paths simultaneously) or temporal (\eg, a two-level system held in a superposition of its energy eigenstates) superposition states. In other words, coherence allows a sensor to behave as a controlled combination of two or more states, enabling a weak external signal to produce a measurable change in the phase or population of these states.
    
    \item \textbf{Quantum entanglement}: Measurement strategies that use entanglement to surpass classical precision limits.
\end{enumerate}

Among these categories, entanglement-enabled sensing represents the strongest route toward \textit{quantum advantage}. However, for near-term consumer electronics, the most practical opportunities arise primarily from the first two categories: quantized systems and controlled quantum coherence. We therefore focus on these two categories to explain the mechanisms underlying deployable sensing architectures for consumer-grade devices. In practical terms, most sensors considered in this paper operate by preparing a known quantum state, allowing an external signal to modify that state, and converting the resulting change into a conventional electrical or optical output.


{However, before introducing a system-level integration workflow, it is useful to identify the four attributes that characterize a widely used class of coherently controlled quantum sensors:}
\begin{enumerate}
    \item \textbf{Discrete energy structure}: One of the fundamental attributes in quantum-mechanical systems is possessing discrete energy levels. In the simplest case of a two-level system, these levels correspond to the ground state $\ket{0}$ and the excited state $\ket{1}$, which are separated by a transition energy. These states serve as two distinguishable sensor states whose energy separation changes in response to an external quantity.
    
    \item \textbf{Initialization and readout}: Before measurement begins, the sensor must be initialized in a known reference state, typically denoted by $\ket{0}$. Then, during the sensing interval, the sensor interacts with the physical signal and evolves into a final state that encodes information about the signal. Afterward, the readout process maps the final quantum state into a corresponding physical quantity, such as fluorescence intensity, optical transmission, or an electrical signal. More general quantum measurements are mathematically described using positive operator-valued measures (POVMs)~\cite{nielsen2010quantum}. For practical purposes, these measures assign probabilities to all the possible sensor outputs.
    
    \item \textbf{Coherent control}: The sensor should permit controlled state manipulation through external fields or pulses. Such control signals prepare the sensor in a state that is maximally responsive to the quantity being measured, analogous to biasing a conventional sensor at its most sensitive operating point.
    
    \item \textbf{Signal coupling}: The sensor must couple to the physical quantity of interest, denoted here by $V(t)$. This interaction is commonly represented by a signal-dependent Hamiltonian, which mathematically describes how the external quantity impacts the sensor's energy and time evolution. The interaction is characterized by a transduction parameter~\cite{QS_Degen}:
    \[
    \gamma = \frac{\partial^q E}{\partial V^q}
    \]
    $\gamma$ relates the changes in the transition energy $E$ to the changes in the external parameter $V(t)$. Typically, the order of operation, $q$ is $1$ for linear interactions, such as the Zeeman effect~\cite{QS_Degen}, whereas $q=2$ for quadratic interactions, such as the second-order Stark effect~\cite{QS_Degen}.

\end{enumerate}

{These four attributes, adapted from the framework proposed by Degen}~\etal~\cite{QS_Degen}{, define the functional chain of coherently controlled sensing. This coherent control process includes a suitable sensing element, a preparation and control method, coupling to the target signal, and a practical readout mechanism. Individual platforms nevertheless impose their own requirements. For instance, spectroscopy-based, photoluminescence-based, and photon-correlation measurements rely on distinct excitation and readout pathways, as discussed in Sections III--V. Even then, signal coupling remains central across these modalities because it determines how the target quantity is converted into an observable optical or electrical response.}

Sensitivity in quantum sensing is defined as the minimum detectable signal per unit time:
\begin{equation} \label{eqn: sensitivity}
v_{\min} \propto \frac{1}{\gamma \sqrt{T_\chi}}.
\end{equation}

As shown in Eq.~\eqref{eqn: sensitivity}, sensitivity ($v_{\text{min}}$) is inversely proportional to the coupling constant, $\gamma$, and to the square root of the decoherence or relaxation time, $T_\chi$. Thus, improved sensing performance (less $v_{\text{min}}$) results when the external signal produces a greater change in the quantum state and the state remains stable over a longer measurement interval. In practice, however, every quantum system is subject to decoherence, and it gradually loses the controlled quantum behavior due to noise and environmental interactions.


\begin{table*}[t]
\centering
\caption{Representative performance and implementation tradeoffs of quantum sensing platforms* relevant to consumer electronics.}
\label{tab:platform_comparison}

\small
\renewcommand{\arraystretch}{1.45}
\setlength{\tabcolsep}{5pt}

\resizebox{\textwidth}{!}{
\begin{tabular}{
>{\raggedright\arraybackslash}p{1.7cm}
>{\raggedright\arraybackslash}p{3.0cm}
>{\raggedright\arraybackslash}p{2.1cm}
>{\raggedright\arraybackslash}p{2.2cm}
>{\raggedright\arraybackslash}p{2.3cm}
>{\raggedright\arraybackslash}p{2.0cm}
>{\raggedright\arraybackslash}p{2.5cm}
>{\raggedright\arraybackslash}p{2.7cm}}
\toprule

\rowcolor{blue!16}
\textbf{Platform} &
\textbf{Representative sensitivity and status} &
\textbf{Power requirements} &
\textbf{Form factor} &
\textbf{Technology readiness} &
\textbf{Cooling} &
\textbf{Integration requirements} &
\textbf{Near-term potential} \\

\midrule

\rowcolor{white}
\textbf{NV centers} &
\textit{{Projected}}: $\sim250\,\mathrm{aT}/\sqrt{\mathrm{Hz}}/\mathrm{cm}^{3}$ (ensemble magnetometry)~\cite{taylor2008high} &
Optical excitation, fluorescence readout, and microwave control~\cite{schirhagl2014nitrogen,hong2013nanoscale} &
Solid-state sensing element; chip- or module-scale implementations~\cite{schirhagl2014nitrogen,NV_SiC_RF} &
Early prototype; room-temperature RF sensing demonstrated~\cite{NV_SiC_RF} &
No cryogenic cooling required~\cite{schirhagl2014nitrogen,hong2013nanoscale} &
Optics, microwave delivery, timing, and readout electronics~\cite{NV_SiC_RF} &
Compact magnetic sensing, RF receivers, and biomedical sensing modules~\cite{schirhagl2014nitrogen,NV_SiC_RF} \\

\midrule

\rowcolor{blue!6}
\textbf{Atomic vapor cells} &
\textit{{Experimentally demonstrated}}: $160\,\mathrm{aT}/\sqrt{\mathrm{Hz}}$ (SERF magnetometry)~\cite{dang2010ultrahigh} &
Laser source, photodetector, heater, and control electronics~\cite{shah2007subpicotesla,liew2004microfabricated} &
Microfabricated vapor cells; compact sensor modules~\cite{liew2004microfabricated,shah2007subpicotesla} &
Near-term product candidate; successfully demonstrated in field settings~\cite{shah2007subpicotesla,Muradoglu:2025blh} &
No cryogenic cooling required; thermal control may be necessary~\cite{liew2004microfabricated,shah2007subpicotesla} &
Optical alignment, cell heating, magnetic shielding, and calibration~\cite{shah2007subpicotesla,Muradoglu:2025blh} &
Portable magnetometry, navigation, timing, and inertial sensing~\cite{Muradoglu:2025blh,shah2007subpicotesla} \\

\midrule

\rowcolor{white}
\textbf{Rydberg atomic sensors} &
\textit{{Experimentally demonstrated}}: $\sim55\,\mathrm{nV\,cm^{-1}/\sqrt{Hz}}$ (RF electric-field sensing)~\cite{jing2020atomic} &
Probe and coupling lasers, optical detection, and frequency-control electronics~\cite{fancher2021rydberg} &
Micrometer-scale vapor cells; complete systems remain at the module or benchtop scale~\cite{kubler2010coherent,fancher2021rydberg} &
Laboratory demonstrations~\cite{fancher2021rydberg,gong2025rydberg,gong2026rydberg,zhang2026rydberg} &
No cryogenic cooling required; cell heating may be necessary~\cite{fancher2021rydberg,kubler2010coherent} &
Laser stabilization, vapor-cell packaging, spectroscopy, and optical readout~\cite{fancher2021rydberg,gong2025rydberg,gong2026rydberg,zhang2026rydberg} &
RF receivers, spectrum monitoring, and wireless/IoT gateways~\cite{fancher2021rydberg,gong2025rydberg,gong2026rydberg} \\

\midrule

\rowcolor{blue!6}
\textbf{Photonic quantum sensors} &
\textit{{Task-dependent}}: determined by the source, detector, optical loss, and sensing protocol~\cite{pirandola2018advancesinPhotonicQS} &
Dependent on the photon source, detector, and optical processing architecture~\cite{pirandola2018advancesinPhotonicQS} &
From chip-scale photonic components to laboratory optical assemblies~\cite{pirandola2018advancesinPhotonicQS,cameron2024adaptive} &
Laboratory demonstrations~\cite{pirandola2018advancesinPhotonicQS,cameron2024adaptive} &
Mostly compatible with room-temperature operation; some single-photon detectors require cooling~\cite{pirandola2018advancesinPhotonicQS} &
Photon source and detector integration, optical alignment, loss management, and correlation processing~\cite{pirandola2018advancesinPhotonicQS,cameron2024adaptive} &
Imaging, biomedical diagnostics, depth sensing, and optical calibration~\cite{pirandola2018advancesinPhotonicQS,cameron2024adaptive} \\

\bottomrule
\end{tabular}
}

\vspace{0.5mm}
\footnotesize {\textit{Note: Sensitivity values correspond to different measured physical quantities and experimental conditions and should therefore \\be interpreted within their respective application domains.} {\textit{Projected sensitivity values are analytical estimates, whereas \\experimentally demonstrated values were measured under the cited operating conditions. \\ *Other platforms like quantum dots, atomic clocks, and trapped ion quantum sensors are also used in consumer electronics applications.}}}
\end{table*}


\section{$\mathrm{III}$ Quantum Sensing Platforms}\label{sec: SensingPlatforms}

In this section, we briefly review the four quantum sensing platforms of particular relevance to near-term consumer electronics: NV centers, atomic vapor cells, Rydberg atomic sensors, and photonic quantum sensors. {The broader family of quantum sensing modalities and device classes, including nuclear magnetic resonance (NMR), atomic clocks, and interferometry, is summarized in Fig.}~\ref{fig: overview}. In the following subsections, we examine, for each platform, the operating principles, the system-level bottlenecks that may limit its deployment, and the potential quantum-mechanical advantage it offers over the classical front ends.

\subsection{Nitrogen-vacancy (NV) centers}
NV centers~\cite{schirhagl2014nitrogen,hong2013nanoscale} are atomic-scale defects in diamond, each consisting of a substitutional nitrogen atom adjacent to a missing carbon atom, that behaves as a solid-state spin sensor. The electronic ground state is a spin triplet, whose sublevels, $\ket{m_s=0}$ and $\ket{m_s=\pm1}$ are separated by a zero-field splitting of $2.87\,\text{GHz}$~\cite{hong2013nanoscale}. When illuminated by an off-resonant green laser, an NV center emits red fluorescence. However, when optically excited from the $\ket{m_s=\pm1}$ state, the NV center preferentially decays through metastable singlet states into the $\ket{m_s=0}$ state, producing a reduced fluorescence yield. This spin-dependent fluorescence contrast is exploited in optically detected magnetic resonance (ODMR) to infer the spin-state populations. {NV center sensors are optically initialized in the} $\ket{m_s=0}$ {state and read out~\via spin-dependent fluorescence, while resonant microwave fields coherently drive transitions between the} $\ket{m_s=0}$ and $\ket{m_s=\pm1}$ {states. A key practical advantage is that initialization, control, and readout can all be performed at room temperature on a solid-state platform without cryogenic cooling.}

The estimated sensitivity of diamond-based NV magnetometers is approximately $250\,\text{aT}/\sqrt{\text{Hz}}/\text{cm}^3$~\cite{taylor2008high}.
This level of sensitivity makes NV centers attractive for compact magnetic sensing, local field mapping, and integrated sensing modules in portable electronics.

From an engineering perspective, NV center sensors are currently at the early prototype stage. Jiang~\etal~\cite{NV_SiC_RF} have demonstrated room-temperature RF sensing using NV centers in SiC. Such solid-state architecture supports chip- or module-scale implementations. However, practical integration still requires optical excitation and fluorescence collection, microwave delivery, precise timing, and readout electronics. Given their current maturity stage and the integration requirements, compact magnetic sensors, RF receivers, and specialized biomedical sensing modules are the most plausible near-term applications~\cite{schirhagl2014nitrogen,NV_SiC_RF}.

\subsection{Atomic vapor cells}
Quantum sensors, based on atomic vapor cells, usually confine alkali atoms, such as rubidium ($^{87}\text{Rb}$), within a small sealed glass enclosure. These atoms can then be optically pumped into a known spin state under laser illumination. In the presence of a magnetic field, their spins then precess at the Larmor frequency, $\omega_L=\gamma B$, where $\gamma$ is the gyromagnetic ratio and $B$ the unknown magnetic field. This precession alters the transmission or the polarization rotation of the probe light, from which the field can be estimated. In the spin-exchange-relaxation-free (SERF) regime, spin-exchange collisions no longer dominate the relaxation process, allowing coherence times of up to $10\,\text{ms}$~\cite{shah2007subpicotesla} and correspondingly high sensitivity. 

These sensors are particularly attractive because they can be microfabricated~\cite{liew2004microfabricated}, require no cryogenic cooling, and can operate using low-power optics and compact heaters. Despite their relative simplicity, atomic vapor cell magnetometers can achieve a sensitivity of approximately $100\,\text{aT}/\sqrt{\text{Hz}}$~\cite{dang2010ultrahigh}. This sensitivity is comparable to that of the superconducting quantum interference devices (SQUIDs), which are among the most sensitive magnetometer technologies reported to date, yet it is achieved without cryogenic cooling. Atomic vapor cells are therefore promising candidates for portable magnetometers and, by extension, compact timing and inertial sensing modules.

From an engineering perspective, atomic vapor cell sensors are among the most mature quantum sensing platforms. They represent a near-term product candidate, as evidenced by several successful field demonstrations~\cite{shah2007subpicotesla,Muradoglu:2025blh}. Such microfabricated cells are favored in compact sensing modules. However, their practical deployment still faces several integration challenges, including optical alignment, controlled cell heating, magnetic shielding, and calibration. Collectively, these requirements result in moderate power consumption and integration complexity. These characteristics make portable magnetometry, navigation, timing, and inertial sensing the most realistic near-term applications for atomic vapor cells~\cite{liew2004microfabricated,shah2007subpicotesla,Muradoglu:2025blh}.

\subsection{Rydberg atomic sensors} 
Rydberg atomic sensors exploit highly excited atomic states in which the outermost electron is only weakly bound. As a result, the atom becomes extremely sensitive to external electric fields. In a typical implementation, two lasers drive the atom through a ladder transition, such as $5S \rightarrow 5P \rightarrow nl$, creating an electromagnetically induced transparency (EIT) feature~\cite{fancher2021rydberg}. EIT is a quantum interference effect in which a coupling laser opens a narrow transmission window within an otherwise absorbing atomic transition. An incident RF or microwave field then couples the nearby Rydberg states, which modifies the overall optical spectrum; in particular, the EIT line splits due to the Autler--Townes effect, and the magnitude of the splitting is proportional to the field strength.

The primary system-level appeal is that the atomic medium responds directly to the RF field, converting it into an optical signal. This capability provides a path toward compact field receivers, spectrum monitors, and RF sensing modules with reduced dependence on conventional front-end components such as antennas, mixers, and low-noise amplifiers~\cite{fancher2021rydberg,kubler2010coherent}. Encouragingly, coherent Rydberg excitation has been demonstrated in micrometer-scale vapor cells when the atomic state and cell material are carefully selected, indicating a reliable route to miniaturization and integration~\cite{kubler2010coherent}.

Facon~\etal~\cite{facon2016sensitive} showed that Rydberg atoms can approach the fundamental \textit{Heisenberg limit} through nonclassical evolution. By evolving through Schr\"odinger cat states, they achieved a sensitivity of $1.2\,\text{mV}\,\text{cm}^{-1}$ over an interaction time of $100\,\text{ns}$. Jing~\etal~\cite{jing2020atomic} experimentally demonstrated an RF electric-field sensitivity of $\sim55\,\mathrm{nV\,cm^{-1}/\sqrt{Hz}}$.

From an engineering perspective, Rydberg atomic sensors remain primarily at the laboratory-demonstration stage, although considerable progress has been made toward micrometer-scale vapor cells and room-temperature operation~\cite{kubler2010coherent,fancher2021rydberg,gong2025rydberg,gong2026rydberg,zhang2026rydberg}. 
Complete systems generally remain at the module or benchtop scale since they require stabilized probe and coupling lasers, vapor-cell packaging, spectroscopic detection, and optical readout.

\subsection{Photonic quantum sensors} 
Photonic quantum sensors use engineered quantum states of light, such as single photons, squeezed light, or entangled beams, to probe a target~\cite{pirandola2018advancesinPhotonicQS}. The measured quantity may be displacement, refractive index, or a phase change. As the measurement is encoded directly in the quantum state of light, the photonic sensors can operate effectively at low photon numbers and offer better noise performance than classical optical schemes in selected regimes~\cite{pirandola2018advancesinPhotonicQS}. In consumer electronics, they could enhance optical subsystems used for imaging, depth sensing, LiDAR, and biomedical applications. Their primary advantage is that photonic quantum protocols can improve weak-signal detection and robustness to noise in power-constrained settings.

Photonic quantum sensors remain largely at the laboratory-demonstration stage, with implementations ranging from chip-scale photonic components to complete optical assemblies~\cite{pirandola2018advancesinPhotonicQS,cameron2024adaptive}. Practical integration requires efficient photon sources and detectors, optical-loss management, precise alignment, and correlation-processing electronics. In addition, some single-photon detectors still require cooling. The most plausible near-term applications include specialized imaging, biomedical diagnostics, depth sensing, and optical calibration~\cite{pirandola2018advancesinPhotonicQS,cameron2024adaptive}. 

Beyond these four, several other platforms show promising potential in consumer electronics applications, including atomic clocks~\cite{QS_Degen}, trapped ions~\cite{QS_Degen}, and quantum dots (QDs)~\cite{chern2019sensing,kim2017FullyStretchableOptoelectricS}. 
Table~\ref{tab:platform_comparison} consolidates the preceding platform-level discussion by comparing representative sensitivities, power requirements, form factors, cooling requirements, integration complexity, technology readiness, and near-term potential across the four platforms. The table is intended to capture the principal system-level tradeoffs; the reported sensitivity values correspond to different physical quantities and operating conditions and should be interpreted within their respective application domains. 

Table~\ref{tab:platform_comparison} also summarizes the technological maturity of the sensing platforms. Based on recent studies, NV center sensors largely fall into the early prototype category, while atomic vapor cell sensors are near-term product candidates for specialized, compact modules. Rydberg atomic sensors and fully integrated photonic quantum sensing systems, however, remain largely confined to laboratory demonstrations only.


\begin{figure*}[htb]
    \centering
    \includegraphics[width=2\columnwidth]{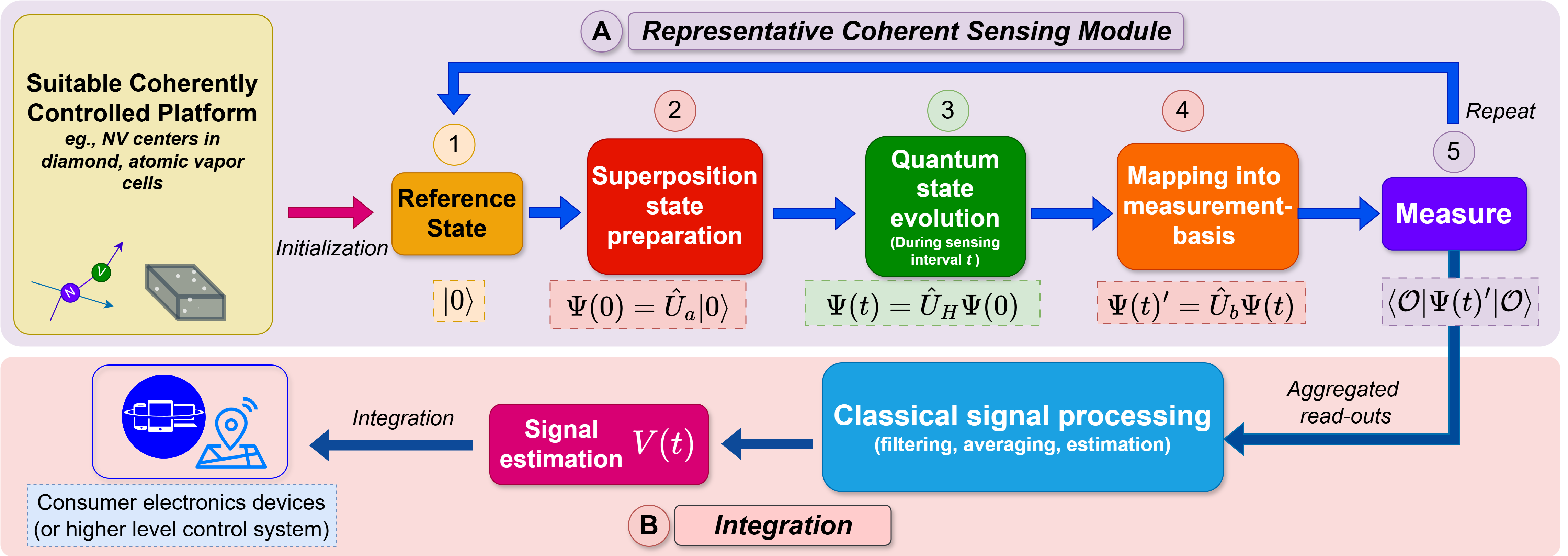}
    
    \caption{
    Representative workflow for integrating a coherently controlled quantum sensor into a consumer electronics system. \footnotesize{\\\textbf{A.} {A Ramsey-type or related pulsed protocol applied to a quantum sensing platform comprises the following steps: the sensor is} initialized (1), prepared in a sensitive superposition state (2), allowed to evolve during the sensing interval $t$ (3), mapped back to the measurement basis (4), and measured repeatedly (5) to estimate the associated signal $V(t)$. The sensing process begins by initializing the sensor in a well-defined reference state ($\ket{0}$). The preparation operation $\hat{U}_a$ (typically a $\frac{\pi}{2}$ pulse) then converts the reference state into a more sensitive superposition state ($\Psi(0)$). During the sensing interval (3), the quantum state evolves under both the sensor's internal dynamics and the external signal, which are jointly represented by $\hat{U}_{H}$. After this interaction, a second control operation $\hat{U}_b$ converts the accumulated change in the quantum state into a measurable population or optical signal, which is then read out in a suitable basis. In contrast, spectroscopy-, photoluminescence-, and photon-correlation-based sensors employ platform-specific excitation, interaction, and readout mechanisms.
    \textbf{B.} {The resulting readouts are aggregated and processed classically} to estimate the signal $V(t)$. This estimation is performed by determining the change from $\Psi(0)$ to $\Psi(t)'$ through classical signal post-processing before integration with a consumer device or higher-level control system.}
    }
    \label{fig: pipeline}
\end{figure*}


\section{$\mathrm{IV}$ A Representative Coherent Sensing Pipeline}
\label{sec: Pipeline}

{We now present a representative pipeline for coherently controlled quantum sensing, building on the principles and platforms introduced above. The sequence shown in Fig.}~\ref{fig: pipeline} {applies directly to Ramsey-type and related pulsed sensing protocols, in which a quantum state is initialized, prepared in a sensitive superposition, allowed to evolve under the target signal, mapped to a measurement basis, and repeatedly read out. An overview of this coherently controlled quantum sensing pipeline also serves as a reference against which the other specialized sensing workflows can be compared.}

The sensing process begins by selecting an appropriate platform based on the target quantity, operating environment, and system constraints. This choice determines both the achievable sensitivity and the complexity of system integration.

The sensor is then initialized in a known reference state, typically denoted as $\ket{0}$. A controlled operation (e.g., $\hat{U}_a$, typically a $\frac{\pi}{2}$ pulse) transforms this reference state into a superposition state that strongly responds to small perturbations.

In the next step, the sensor undergoes controlled evolution. During this phase, the quantum state evolves under the combined effect of the internal and the signal Hamiltonians, while interacting with the external environment. This process allows the external parameter of interest, $V(t)$, to be encoded into the state as the phase or amplitude information.

After the interaction, a second control operation, $\hat{U}_b$, typically the inverse of the initial pulse, $\hat{U}_b=\hat{U}^{\dagger}_a$, maps the system onto a preferred measurement basis, and the sensor is then read out. As quantum measurements are probabilistic, repeated initialization-to-measurement cycles are required to reliably estimate the signal. The resulting data are then processed classically through averaging, filtering, and calibration to recover $V(t)$, while accounting for noise and device-specific imperfections.

Finally, the processed signal is integrated into a higher-level system. At this stage, the quantum sensor functions as a specialized front-end module that provides enhanced measurement capabilities for a conventional consumer electronics application.

{Not all sensing modalities fully follow this coherent-control sequence; several replace parts of it with platform-specific transduction and readout mechanisms. Rydberg atomic receivers commonly use continuous optical driving and infer an RF field from changes in the EIT spectrum or Autler--Townes splitting}~\cite{fancher2021rydberg,gong2025rydberg}{. Quantum dot assays and optoelectronic devices encode the measured quantity in photoluminescence intensity or photocurrent rather than in an explicitly prepared superposition state}~\cite{chern2019sensing,kim2017FullyStretchableOptoelectricS}{. Likewise, the quantum-assisted imaging scheme of Cameron}~\etal~\cite{cameron2024adaptive} {extracts aberration information from spatial correlations between photon pairs. These approaches, therefore, do not require the explicit superposition preparation and inverse basis mapping steps, shown in Fig.}~\ref{fig: pipeline}{, yet, they share the other downstream functions, including repeated or aggregated acquisition, classical processing, signal estimation, and integration with a higher-level system.}

\section{$\mathrm{V}$ Applications in Consumer Electronics}\label{sec: applications}
Quantum sensing is no longer confined to fundamental physics. Recent work has demonstrated promising routes to integrating quantum sensing technologies with hardware relevant to consumer electronics, with applications reported in precision navigation~\cite{Muradoglu:2025blh}, healthcare diagnostics and wearable devices~\cite{chern2019sensing,kim2017FullyStretchableOptoelectricS}, RF reception~\cite{NV_SiC_RF,gong2025rydberg,gong2026rydberg}, and adaptive imaging~\cite{cameron2024adaptive}. This section presents representative case studies from these application areas, emphasizing their key advantages and principal integration bottlenecks. Although some of the examples discussed below have not yet reached commercial consumer products, they demonstrate sensing capabilities that could support future consumer platforms provided that continued progress is made in miniaturization, energy efficiency, manufacturing cost minimization, and system integration.

\subsection{Navigation and Positioning} 
{Atomic magnetometry has been demonstrated as a basis for magnetic-anomaly navigation in environments where the Global Positioning System is unavailable}~\cite{Muradoglu:2025blh}. Murado\u{g}lu~\etal~\cite{Muradoglu:2025blh} presented a navigation system built around a quantum magnetometer integrated into a practical navigation front end. The sensing element in this system is a scalar, optically pumped rubidium magnetometer that detects variations in Earth’s magnetic field~\via atomic spin precession in a vapor cell. The light sources, photodetectors, and cell heaters are integrated into a compact sensor head, and a classical vector magnetometer and an inertial navigation system (INS) complement the quantum sensor to form a complete magnetic-anomaly-navigation stack.

Crustal magnetic anomalies are typically only on the order of $10$--$100\,\text{nT}$ over distances of several kilometers, whereas magnetic interference from the host platform can be $100$--$1000$ times larger. Resolving these weak signatures, therefore, requires both high sensitivity and long-term stability. The reported hardware achieves a sensitivity below $80\,\mathrm{fT}/\sqrt{\mathrm{Hz}}$ at Earth’s magnetic field strength, a bandwidth of $250\,\text{Hz}$, a mass of approximately $70\,\text{g}$, a volume of $144\,\text{cm}^3$, and a power consumption below $15\,\text{W}$. These characteristics make the sensor suitable for integration into drones and aircraft, demonstrating progress beyond laboratory-scale operation toward deployable navigation systems.

The coordination between the sensing hardware and system software is central to the practical relevance of this work. Rather than treating the quantum magnetometer as a stand-alone instrument, the system fuses its output in real time with vector magnetic measurements, INS estimates, velocity data, and magnetic-anomaly maps. A denoising and map-matching engine learns the vehicle’s magnetic signature online, without dedicated calibration maneuvers. It is this system-level integration that translates the magnetometer’s raw sensitivity into bounded navigation error.

In airborne and ground trials, the system consistently outperformed a strategic-grade INS, reducing positioning errors by up to a factor of 46 and achieving a best final positioning error of $22\,\text{m}$, while remaining sufficiently compact for practical avionics packaging. The broader lesson for the consumer and mobility applications is that the value of quantum sensing arises not only from high sensitivity but also from the careful co-design of the sensor, its placement, and the surrounding estimation stack.

Although the reported demonstration targets airborne navigation, the underlying sensing approach is likely to reach consumer-adjacent platforms, such as drones, autonomous robots, and vehicles operating in GPS-denied environments. Earlier demonstrations of millimeter-scale, low-power atomic magnetometers indicate a technically credible path toward more compact sensing modules~\cite{schwindt2004chip,shah2007subpicotesla}. With further miniaturization of the optical and control components, these sensors could eventually complement conventional microelectromechanical systems (MEMS), camera-based, and wireless positioning sensors in portable or infrastructure-assisted consumer devices.


\begin{table*}[t]
\caption{{Recent works on quantum and quantum-enabled sensing applications relevant to consumer electronics.}}
\label{tab: case_studies}

\centering
\small
\renewcommand{\arraystretch}{1.45}
\setlength{\tabcolsep}{5pt}

\resizebox{\textwidth}{!}{
\begin{tabular}{
>{\raggedright\arraybackslash}p{2.8cm}
>{\raggedright\arraybackslash}p{2.6cm}
>{\raggedright\arraybackslash}p{2.7cm}
>{\raggedright\arraybackslash}p{3.2cm}
>{\raggedright\arraybackslash}p{4.0cm}}
\toprule

\rowcolor{blue!15}
\textbf{Authors \& year} &
\textbf{Application domain} &
\textbf{Platform} &
\textbf{Maturity stage} &
\textbf{Main integration bottleneck} \\

\midrule

\rowcolor{blue!6}
\makecell[l]{Muradoğlu~\etal\cite{Muradoglu:2025blh}\\2025} &
\makecell[l]{Navigation\\\& Positioning} &
\makecell[l]{\textit{Atomic vapor}\\cell} &
\makecell[l]{Field-validated} &
\makecell[l]{Magnetic interference \\ suppression and map \\dependence} \\

\rowcolor{white}
\makecell[l]{Chern~\etal\cite{chern2019sensing}\\2019} &
Healthcare &
\makecell[l]{\textit{QD}-based fluorescence\\assays} &
\makecell[l]{Application\\\textit{(Smartphone-readable}\\\textit{preclinical demonstrations)}} &
Chemistry and calibration \\

\rowcolor{blue!6}
\makecell[l]{Kim~\etal\cite{kim2017FullyStretchableOptoelectricS}\\2017} &
\makecell[l]{Healthcare\\(Wearable)} &
\makecell[l]{Stretchable\\\textit{QD} optoelectronics} &
\makecell[l]{Application\\\textit{(Patch-level}\\\textit{PPG demonstrated)}} &
Manufacturing and long-term stability \\

\rowcolor{white}
\makecell[l]{Jiang~\etal\cite{NV_SiC_RF}\\2023} &
RF / IoT &
\makecell[l]{\textit{NV centers} in SiC} &
\makecell[l]{Laboratory POC\\\textit{(Room-temperature}\\\textit{RF sensing demonstrated)}} &
Coherence and system integration \\

\rowcolor{blue!6}
\makecell[l]{Gong~\etal\cite{gong2025rydberg}\\2025} &
RF / IoT &
\makecell[l]{\textit{Rydberg} atomic receiver} &
\makecell[l]{Laboratory POC\\\textit{(Lab-scale atomic}\\\textit{receiver demonstrated)}} &
Laser and vapor-cell packaging \\

\rowcolor{white}
\makecell[l]{Cameron~\etal\cite{cameron2024adaptive}\\2024} &
Imaging &
\makecell[l]{Entangled \textit{photon}s} &
\makecell[l]{Laboratory POC} &
Long acquisition time \\

\bottomrule
\end{tabular}
}

\vspace{0.5mm}
\footnotesize{\textit{Note: Representative quantum sensing demonstrations in consumer-electronics-related applications. The maturity stage of the technology is classified into three standardized categories: laboratory proof of concept (POC), application prototype, and field-validated prototype.}}
\end{table*}

\subsection{Health and Wearables}

{Healthcare and wearable technologies illustrate a related category based on quantum dots (QDs), nanoscale semiconductor structures whose size-tunable optical properties arise from quantum confinement}~\cite{kouwenhoven2001fewQuantumDot}. {QDs are therefore quantum-enabled sensing materials rather than coherently controlled quantum sensors as discussed in Section-IV.} Chern~\etal~\cite{chern2019sensing} reviewed smartphone-readable QD-based fluorescence-linked immunosorbent assays (FLISAs) for home diagnostics, where QDs serve as fluorescent labels that enable compact, smartphone-based readout. Their principal advantages include high brightness, strong photostability, narrow emission spectra, and broad absorption bands. The broad absorption bands, in particular, enable the detection of multiple emission colors with a single excitation source, thereby significantly reducing equipment costs and complexity.

Experimental results indicate strong potential for point-of-care diagnostics. QD-based fluorescence-linked immunosorbent assays have achieved limits of detection below $5\,\text{pg/mL}$, compared with $20\,\text{pg/mL}$ for conventional colorimetric assays, and QD-based C-reactive protein (CRP) tests have also demonstrated shorter assay times of approximately $50\,\text{min}$. Smartphone-based fluorescence-imaging systems have likewise detected lower analyte concentrations than systems using conventional dyes~\cite{chern2019sensing}. Practical deployment, however, remains constrained by assay chemistry, calibration requirements, and the need for reliable, well-controlled optical readout.

Kim~\etal~\cite{kim2017FullyStretchableOptoelectricS} demonstrated a stretchable optoelectronic sensor based on colloidal QDs for continuous physiological monitoring. The system integrates QD light-emitting diodes (QD-LEDs) and QD photodetectors with graphene electrodes on a prestrained elastomer substrate to form a thin, flexible patch. The device measures photoplethysmography (PPG) signals by illuminating the skin and detecting variations in the reflected or transmitted light associated with changes in blood volume.

The prototype maintained reliable operation under mechanical deformation, demonstrating stretching of up to $70\%$ and a bending radius as small as $35\,\mu\text{m}$. This mechanical compliance enables stable sensing on moving skin and supports its use as an adhesive patch on sites such as the ear or finger. Although large-scale manufacturing, device uniformity, and calibration remain challenging, QD-based optoelectronics offer clear advantages for wearable sensing owing to their tunable emission wavelengths, narrow spectral bandwidths, and compatibility with diverse substrates.

\subsection{RF, Wireless, and IoT}

A practical entry point for quantum sensing in RF and Internet of Things (IoT) applications is the integration with existing wireless infrastructure. Jiang~\etal~\cite{NV_SiC_RF} demonstrated room-temperature RF sensing using NV centers in silicon carbide (SiC). Compared with diamond, SiC offers greater compatibility with wafer-scale fabrication and conventional electronic-device processing while retaining optically addressable spin defects. In the demonstrated platform, an ensemble of NV centers is optically excited using a $980\,\text{nm}$ laser and controlled through microwave pulse sequences to detect RF signals.

A primary engineering challenge was the short coherence time. The authors applied XY8 dynamical decoupling sequences, extending the coherence time from approximately $2.07\,\mu\text{s}$ under a spin-echo sequence to $28.1\,\mu\text{s}$, an improvement of more than one order of magnitude. The resulting longer interrogation window enabled correlation spectroscopy of an RF signal near $900\,\text{kHz}$, resolving a peak at $898\,\text{kHz}$ with a spectral resolution of $10\,\text{kHz}$. Synchronized readout further improved the resolution to $0.01\,\text{kHz}$ for a signal at $899.28\,\text{kHz}$~\cite{NV_SiC_RF}.

The significance of this work lies not only in the sensing defect but also in the combination of a manufacturable host material, pulse-sequence engineering, and high-resolution readout. This combination makes SiC a credible solid-state platform for compact quantum RF sensors. Practical deployment, however, will require greater coherence, fewer control errors, and the integration of optical excitation, microwave delivery, timing, and readout electronics into a stable, low-cost package.

Gong~\etal~\cite{gong2025rydberg} investigated Rydberg-atom quantum receivers (RAQRs), which use an optically driven atomic vapor cell in place of portions of a conventional RF receiver front end. An incident RF field perturbs the Rydberg energy levels, and the resulting response is read out optically through electromagnetically induced transparency (EIT) and Autler--Townes splitting. In a superheterodyne configuration, an additional local-oscillator field mixes the signal within the atomic medium itself, producing an intermediate-frequency or baseband output.

The principal advantage is architectural: RF-to-optical transduction and frequency conversion occur directly within the sensing medium, potentially reducing dependence on conventional electronic mixer chains. This makes the approach relevant to compact Wi-Fi and other wireless receivers. Practical deployment, however, remains constrained by the need to miniaturize and stabilize the lasers, vapor cell, optical readout, and frequency-control components while reducing total power consumption and cost.

\subsection{Imaging}
Emerging quantum imaging techniques can outperform the conventional methods in specific operating regimes or provide capabilities that are otherwise difficult to achieve. Cameron~\etal~\cite{cameron2024adaptive} proposed a quantum-assisted adaptive-optics (QAO) method for correcting aberrations in label-free microscopy, where the point-spread function (PSF) cannot be measured directly. The method uses spatial correlations between entangled photon pairs to extract aberration information without requiring a guide star or a sample-dependent image-quality metric. In the experimental setup, spatially entangled photon pairs illuminate the sample, and the transmitted photons are detected using a single-photon-sensitive camera. The aberration is corrected by maximizing the central peak of a measured second-order spatial-correlation projection, which is proportional to the PSF. This approach offers an engineering advantage over conventional image-based adaptive optics, whose correction accuracy can depend strongly on the chosen optimization metric and the sample's structure.

The experimental results demonstrate promising aberration correction in both classical and quantum imaging configurations. In bright-field transmission imaging of a biological sample, the structural similarity index measure (SSIM) relative to an aberration-free reference increased from $77.89\%$ before correction to $98.41\%$ after QAO correction. In a defocus-correction experiment involving a three-dimensional object, conventional image-based metrics converged to incorrect solutions, whereas QAO recovered the correct correction and achieved an SSIM of $96.83\%$. The method was also applied to a quantum bright-field imaging configuration, in which aberrations had reduced the signal-to-noise ratio of the anticorrelation image to approximately 3. QAO increased this ratio to 15, recovering the expected spatial detail.

Although the current implementation remains laboratory scale, its relevance to consumer electronics lies in compact aberration correction and optical calibration. Continued advances in CMOS-compatible single-photon avalanche-diode (SPAD) arrays and integrated room-temperature single-photon detectors~\cite{morimoto2020megapixel,govdeli2024room} could support simplified implementations for smartphone-assisted microscopy, home healthcare imaging, and periodic calibration of augmented-reality and virtual-reality (AR/VR) optical systems.

Practical deployment, however, remains constrained by the relatively long acquisition time. With the electron-multiplying charge-coupled-device (EMCCD) camera, a single correlation measurement requires approximately $1$--$3\,\text{min}$, and correcting multiple aberration modes can consequently require several hours. Faster detectors and more efficient acquisition strategies could help mitigate this limitation.

Table~\ref{tab: case_studies} reveals a consistent pattern across the reviewed studies: quantum sensing technologies are being evaluated for consumer-electronics-related applications spanning navigation, healthcare diagnostics, wearable devices, RF/IoT systems, and imaging. The underlying platforms and their functions are diverse: atomic vapor cells support magnetic-anomaly navigation; quantum-dot systems support diagnostic assays and flexible wearable patches; NV centers in SiC provide RF sensing; and Rydberg-atom receivers offer an alternative front end for wireless reception. According to the technology-maturity scale presented in Table~\ref{tab: case_studies}, these studies span development stages ranging from laboratory proof-of-concept demonstrations to application prototypes and field-validated prototypes.

{Among all the application areas reviewed, navigation and positioning represent the most realistic initial commercial entry point, particularly for drones, autonomous robots, and vehicles operating in GPS-denied environments}~\cite{Muradoglu:2025blh}{, rather than for highly constrained handheld devices.} Reliable positioning in such environments offers a system-level benefit, large enough to justify the remaining overhead associated with optical components, thermal control, magnetic shielding, and calibration. Healthcare, RF/IoT, and imaging applications also remain promising~\cite{chern2019sensing,kim2017FullyStretchableOptoelectricS,NV_SiC_RF,gong2025rydberg,cameron2024adaptive}, though their broader deployment requires further progress in manufacturing scalability, system integration, power consumption, or acquisition speed.

\section{$\mathrm{VI}$ Conclusion}\label{sec: discussion}

In this paper, we examined where quantum sensing can offer practical advantages in consumer electronics systems. The reviewed case studies show that these advantages are application-dependent rather than universal: atomic magnetometers enable sensitive magnetic-anomaly navigation; quantum-enabled materials support flexible healthcare sensing; NV centers and Rydberg atoms offer alternative mechanisms for RF detection; and photon correlations enable aberration correction under challenging imaging conditions. Collectively, these results establish sensitivity, noise resilience, and specialized transduction as the principal benefits that may justify the added complexity of quantum hardware. Translating these capabilities into deployable systems nevertheless requires overcoming challenges, including calibration and chemistry control, scalable and reproducible manufacturing, longer coherence times, and compact integration of optical, microwave, thermal-control, and vapor-cell components. 

The technological maturity of the reviewed platforms remains uneven. Atomic vapor-cell sensors currently offer the clearest path toward specialized deployable modules, NV-center systems are progressing through early prototypes, and Rydberg and photonic platforms remain predominantly at the laboratory-demonstration stage. 

Overall, the near-term significance of quantum sensing lies not in replacing complete classical sensing systems but in augmenting them at measurement-critical points. The first commercial deployments are likely to emerge where a single quantum measurement block, such as a chip-scale vapor-cell magnetometer or a SPAD-based correlation detector, can be inserted into an existing classical signal-processing chain with minimal changes to the surrounding system. Progress will be governed less by fundamental sensitivity than by the co-packaging of lasers, vapor cells, and detectors on mass-manufacturable platforms. Estimation software that can recover a stable signal from an imperfect device will also play an important role. Hybrid classical--quantum architectures offer the most realistic pathway for converting platform-level quantum advantages into measurable system-level improvements. Whether those advantages can be reliably delivered at the scale for broader consumer adoption will depend on continued advances in miniaturization, calibration, and manufacturing.

\bibliographystyle{bib_style}
\bibliography{main_bib_file}
\begin{IEEEbiography}{Sounak Bhowmik}{\,}
is currently pursuing the Ph.D. degree in electrical and computer engineering at Southern Methodist University, Dallas, TX, USA. Contact him at sbhowmik@smu.edu.
\end{IEEEbiography}

\begin{IEEEbiography}{Himanshu Thapliyal}{\,}
is the inaugural Walden and Paula Rhines Endowed Professor of Quantum Informatics in the Lyle School of Engineering at Southern Methodist University, Dallas, TX, USA. He is a Senior Member of the IEEE. Contact him at hthapliyal@smu.edu.
\end{IEEEbiography}
\balance
\end{document}